\documentclass[journal=jacsat,manuscript=article]{achemso}

\usepackage[version=3]{mhchem} 
\usepackage{xcolor}

\author{Ankur Singh}
\email{ankur.singh@wisc.edu}
\affiliation[UW-Madison]
{Department of Electrical and Computer Engineering, University of Wisconsin–Madison, Madison, WI, USA}

\author{Akhilesh Jaiswal}
\affiliation[UW-Madison]
{Department of Electrical and Computer Engineering, University of Wisconsin–Madison, Madison, WI, USA}

\title[An \textsf{achemso} demo]
  {OptiLookUp: An Optical ROM-Based Lookup Table Engine for Photonic Accelerators}

\abbreviations{IR,NMR,UV}

\begin{document}

\begin{abstract}
Read-only memory (ROM) provides deterministic access to predefined data mappings. Extending ROM concepts to the optical domain enables high-bandwidth, low-latency, and parallel memory access, but realising compact, reconfigurable optical ROM remains challenging due to loss, wavelength control, and integration constraints. This work presents a high-speed, reconfigurable photonic ROM architecture implemented using integrated microring resonators (MRRs). The ROM encodes predefined input–output mappings directly in the spectral response of the photonic devices, enabling deterministic lookup-based operation without dynamic computation during readout. To improve scalability and reduce cumulative insertion loss, the architecture employs compact banked sub-arrays that are selectively addressed through an optical decoding mechanism. Reconfigurability is achieved using transistor-based optical selectors, allowing different ROM banks to be activated without physical light rerouting or interferometric structures. The proposed photonic ROM is designed and evaluated using device-level simulations based on the GlobalFoundries 45SPCLO silicon photonics platform. Simulation results demonstrate reliable operation at data rates up to 12.5 GHz, with stable light-to-current transfer characteristics obtained through integrated photodiode readout. The optical ROM can be used to implement nonlinear activation functions utilised in photonic accelerator architectures, including sigmoid, tanh, ReLU, and exponential mappings. In this use case, the ROM operates as a lookup table that stores predefined nonlinear transfer characteristics, demonstrating that complex nonlinear behaviours can be realised through static optical responses rather than active photonic computation. The use of activation functions in this work serves to illustrate the flexibility of the proposed ROM framework and does not constrain its general applicability to other lookup-based functions.
\end{abstract}

\section{Introduction}
The rapid growth of artificial intelligence (AI) and machine learning (ML) workloads has imposed extraordinary demands on computing hardware. Training compute for state-of-the-art AI models has grown by more than ten billion-fold since 2010, with a doubling time of approximately six months in the deep learning era~\cite{1sevilla2022compute,
2thompson2020computational}, far outpacing the improvements historically delivered by Moore's Law. This trajectory has created a fundamental energy and throughput wall that electronic processors, including modern graphics processing units (GPUs), increasingly
struggle to overcome~\cite{3shastri2021photonics}. To address this bottleneck, silicon photonics has emerged as a compelling platform, offering speed-of-light signal propagation, low-loss optical interconnects, and the ability to exploit wavelength division multiplexing (WDM) for massively parallel data processing~\cite{4shen2017deep, 5feldmann2021parallel}. Integrated photonic neural network architectures have demonstrated orders-of-magnitude improvements in energy efficiency for linear matrix-vector operations, motivating the construction of complete photonic computing pipelines in monolithic CMOS-compatible silicon photonics processes~\cite{6bandyopadhyay2024single}.

Despite these advances, a critical and often underappreciated bottleneck in both digital and analog photonic computing remains the implementation of nonlinear functions. In the context of deep neural networks, nonlinear activation functions such as sigmoid, hyperbolic tangent (tanh), rectified linear unit (ReLU), and softmax are indispensable for enabling universal approximation capabilities~\cite{7williamson2019reprogrammable, 8miscuglio2018all}. Similarly, scientific and signal-processing workloads require evaluation of transcendental functions including sine, exponential, logarithm, and square root. In electronic
hardware, these functions are handled by dedicated Special Function Units (SFUs) in GPUs configurable pipelined circuits built around minimax polynomial interpolation tables that consume significant energy in mature CMOS nodes~\cite{9oberman2005high, 10li2016sfu}. In the optical domain, the situation is considerably more challenging: passive silicon photonic devices lack gain and operate with a restricted set of physical nonlinearities, making it difficult to engineer arbitrary nonlinear transfer characteristics analytically. Existing proposals for optically implemented activation functions require either all-optical nonlinear media that are difficult to integrate at scale~\cite{11zuo2019all}, or hybrid
electro-optic feedback loops that introduce latency and analog-to-digital conversion overhead~\cite{12xu2022reconfigurable}. A fundamentally different approach one that avoids active photonic computation altogether is therefore highly desirable.

A well-established solution in electronic hardware is the use of lookup tables (LUTs) and read-only memory (ROM) to approximate nonlinear and transcendental functions~\cite{13meher2010optimized, 14xie2020twofold, 15dong2020plac}. In this approach, a pre-computed table of function values is stored in memory and indexed by the input argument, replacing arithmetic logic with a deterministic memory-read operation. This trades area against computational simplicity and delivers predictable, single-cycle latency. Piecewise linear interpolation can further reduce table size while maintaining sufficient approximation accuracy for neural network inference~\cite{15dong2020plac}. The key architectural insight is that any smooth, bounded nonlinear function can be represented to arbitrary precision through tabulation, irrespective of its analytical complexity. Extending this principle to the optical domain offers the prospect of performing function evaluation at photonic speeds with sub-nanosecond access times
and high-bandwidth parallel readout without exploiting any optical nonlinear effects in the device. Although silicon MRRs are physically capable of nonlinear optical behaviour such as the Kerr effect and two-photon absorption at high optical powers, the proposed architecture operates entirely in the linear resonant filtering regime, where the rings simply pass or drop light based on wavelength matching. The nonlinear function
is not generated by any nonlinear optical interaction; rather, it is encoded in the pre-programmed resonance states of the array, with the nonlinearity residing in the stored data rather than in the device physics.

Several prior works have explored the concept of photonic read-only memory, yet each carries significant limitations. The foundational proposal by Barrios and Lipson demonstrated that a silicon microring resonator with a floating-gate charge storage mechanism could encode binary memory states as resonance shifts, producing a digital optical output~\cite{16barrios2006silicon}. Building on this concept, an SOA-based frequency-encoded ROM architecture was subsequently proposed by Mukherjee, encoding address inputs through  ptical frequency states to select stored data words~\cite{17mukherjee2011all}, while further SOA-based ROM designs exploiting cross-gain modulation were introduced for simple logic encoding~\cite{18jung2009design}. More recently, an RSOA-based decoder-driven ROM operating at 125~Gb/s was demonstrated by Bosu and Bhattacharjee~\cite{19bosu2023design}, and Si$_3$N$_4$ microring structures were proposed for truth-table-based binary ROM operation~\cite{20saharia2022proposed}. However, all of these works share significant limitations. First, prior photonic ROM demonstrations produce exclusively digital binary outputs, with no mechanism to generate a continuous analog value suitable for direct use as a neural network activation output. Second, none support reconfiguration without physical device intervention such as laser annealing or charge injection making it impractical to update stored function values in deployed systems. Third, none provide a scalable addressing mechanism capable of selecting among multiple independent banks, constraining array sizes to small proof-of-concept structures incompatible with practical lookup table widths. Fourth, cumulative through-port insertion loss from cascaded non-resonant rings degrades signal integrity as array size grows, with no architectural solution proposed to mitigate this. Finally, no prior work has applied a photonic ROM to neural network activation function lookup, nor demonstrated dual-mode operation in which the same integrated structure delivers both a digital waveguide output and a continuous analog photocurrent.

In this work, we present a high-speed, reconfigurable photonic ROM architecture that overcomes these limitations using integrated microring resonators (MRRs) on the GlobalFoundries 45SPCLO monolithic silicon photonics
platform~\cite{21rakowski202045nm, 22stojanovicetal2018monolithicsilicon}. The architecture encodes predefined input--output mappings directly in the spectral response of MRR banks, enabling deterministic lookup-based photonic operation without dynamic optical computation during readout. Scalability and loss management are addressed through a banked sub-array organisation in which the optical signal is routed only through the selected bank of MRRs via a optical selector, significantly reducing the cumulative through-port insertion loss that would otherwise accumulate from non-resonant rings across a full cascaded array. To support analog operation, the output waveguides of each 4×4 MRR bank are fed into a multi-mode interference (MMI) combiner tree, which merges the optical signals onto a photodetector. Bank selection is governed by a content addressable memory (CAM) cell, which activates only the addressed bank. Alternatively, in digital mode, the same architecture operates without the photodetector stage the output waveguides from each bank are read directly as optical signals, preserving multi-level digital outputs in the optical domain for digital computing applications. This dual-mode capability not demonstrated in any prior photonic ROM substantially broadens the applicability of the architecture beyond binary digital memory to continuous-valued function approximation for photonic neural network accelerators~\cite{23pour2020experimental, 24huang2021silicon}.

Device-level simulations based on the GF45SPCLO process design kit (PDK) confirm reliable operation at data rates up to 12.5~GHz, with stable light-to-current transfer characteristics obtained via integrated photodiode readout. As a representative use case, we demonstrate that the photonic ROM can implement four key nonlinear activation functions ReLU, exponential, sigmoid, and tanh (REST) by storing pre-computed function values as static optical responses in the MRR array. This demonstrates that complex nonlinear behaviours can be realised through static spectral encoding rather than through active photonic computation, establishing the first photonic ROM-based activation function lookup unit on a monolithic silicon photonics platform. The remainder of this paper is organized as follows: Section~2 describes the 4-bit optical ROM; Section~3 presents the banked photonic ROM; Section~4 presents the nonlinear function implementation based on photonic ROM; Section~5 covers performance and comparison; and Section~6 concludes with future directions.

\section{Results and discussion}

\subsection{Optical Read Only Memory (ROM)}

\begin{figure}[t]
    \centering
    \includegraphics[width=1\linewidth]{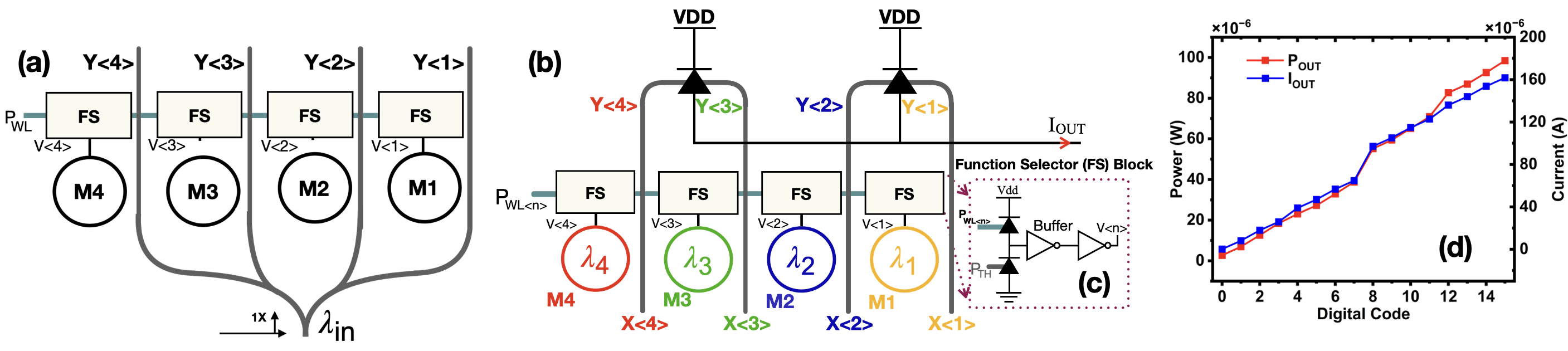}
    \caption{(a) Optical 4-bit ROM implemented using microring resonators. (b) Nonlinear function mapping mechanism enabled by the optical ROM architecture. (c) Function-select unit for ROM operation. (d) Linear mapping of a digital input code to the corresponding output optical power or electrical current.}
    \label{fig:bitweight_mrr}
\end{figure}

The photonic ROM illustrated in Fig.~\ref{fig:bitweight_mrr}(a) implements a 4-bit ROM using microring resonators integrated on a silicon photonic platform. Each microring incorporates an integrated diode, and its electrical drive is provided by the function selection (FS) block, which consists of a balanced photodiode and two cascaded inverters acting as buffer. In this FS block, one photodiode is illuminated by the optical word-line power $P_{\mathrm{WL}}$, while the second photodiode is biased by a reference optical threshold power $P_{\mathrm{TH}}$. The output of the balanced photodiode is connected to a buffer that generates an electrical control voltage $V\langle n\rangle$, where $n=\{1,2,3,4\}$ corresponds to the four microring elements. When the applied word-line optical power satisfies $P_{\mathrm{WL}} > P_{\mathrm{TH}}$, the FS block outputs $V\langle n\rangle = V_{\mathrm{DD}}$, thereby electrically biasing the corresponding microrings. This bias shifts the resonance wavelengths of the microrings to align with the input optical wavelength $\lambda_{\mathrm{in}}$. Conversely, when $P_{\mathrm{WL}} < P_{\mathrm{TH}}$, the FS block outputs $V\langle n\rangle = 0~\mathrm{V}$, leaving the microrings unbiased and detuned from $\lambda_{\mathrm{in}}$.

Microrings that are electrically biased by the FS block resonate at $\lambda_{\mathrm{in}}$ and strongly attenuate the optical signal through resonance-induced coupling, resulting in negligible output power at the corresponding waveguide ports. In contrast, microrings that are not connected to the FS block do not resonate at $\lambda_{\mathrm{in}}$, allowing the optical signal to propagate through the waveguides with minimal attenuation. As shown in Fig.~1(a), microrings $M_2$ and $M_3$ are electrically disconnected from the FS block and are included primarily to preserve the geometrical symmetry of the ROM architecture; these elements do not participate in the electrical biasing and can be disconnected or omitted where required. Accordingly, for the configuration shown in Fig.~1(b), when the optical word-line power $P_{\mathrm{WL}}$ is applied, the photonic ROM outputs correspond to the logic state $Y\langle4\!:\!1\rangle = 0110$.
As shown in Fig.~1(a), the optical ROM operates in a digital mode using a single wavelength with uniform optical power. For analog computation, a photodiode-based readout is employed, as illustrated in Fig.~1(b), to convert the optical signal into an electrical current. To enable this, a weighted optical power scheme is used, where inputs $X\langle1\rangle$–$X\langle4\rangle$ are assigned binary-weighted powers ($1\times$ to $8\times$). The resulting optical outputs are combined and converted by balanced photodiodes into a current proportional to the total weighted input power, as shown in Fig.~1(d) and discussed in the subsequent section.


\subsection{Scalable Banked Photonic ROM Architecture}

\begin{figure}[t]
    \centering
    \includegraphics[width=1\linewidth]{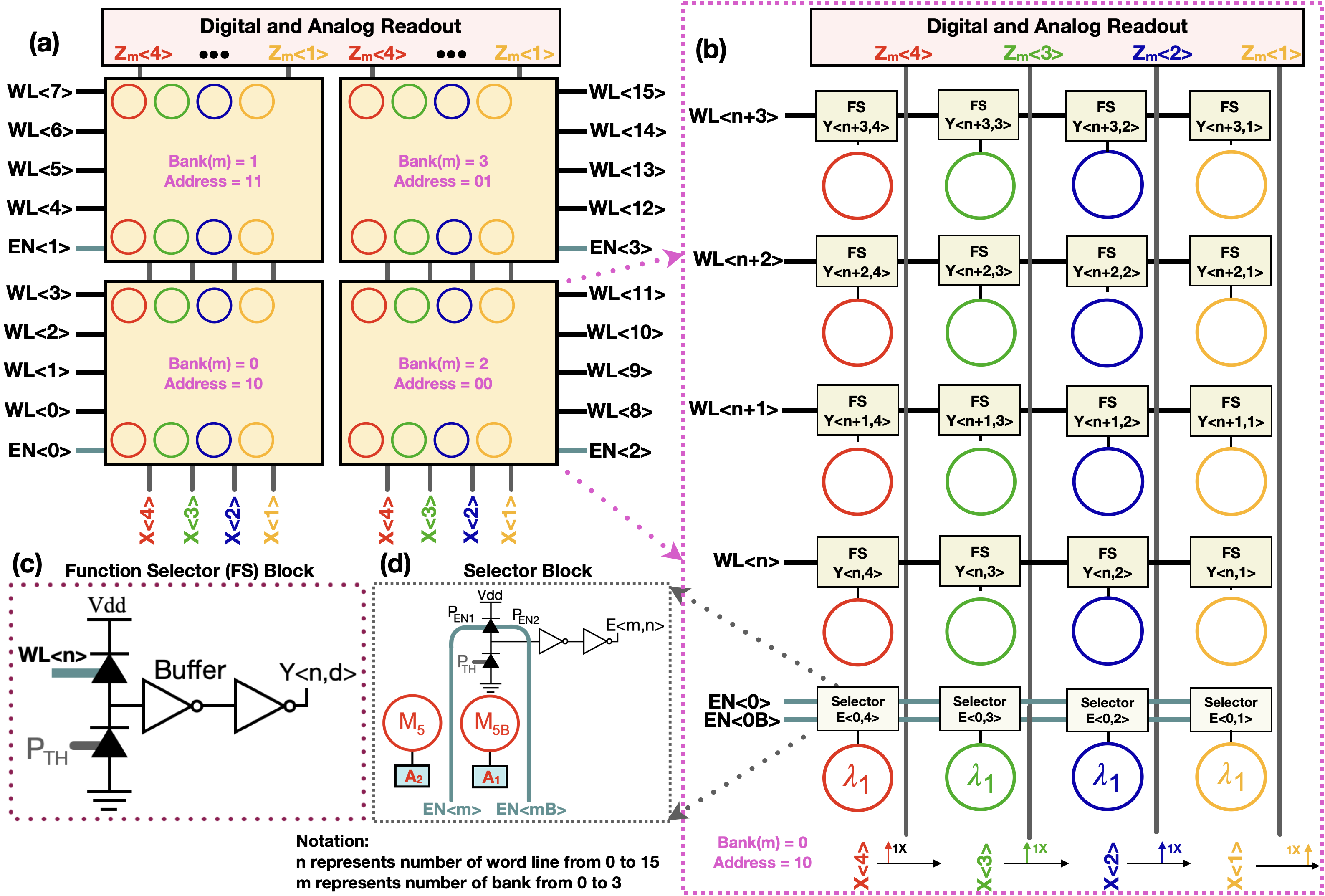}
    \caption{Proposed scalable banked photonic ROM architecture. (a) Overall banked ROM organization showing all four banks for scalable memory expansion. (b) Proposed ROM bank, microrings with and without electrical connections to function-select (FS) blocks. (c) Function-select unit for ROM operation. (d) Optical selector block used to enable the matched ROM bank based on 2-bit MSB input.}
    \label{fig:scalable_rom}
\end{figure}

To scale the photonic ROM capacity while preserving low insertion loss and fast access time, we organize the memory into four independently addressable banks, each implemented as a 5-row × 4-column microring array, as shown in Fig.~\ref{fig:scalable_rom}(b). This banked organization is mainly intended to limit the optical loss that accumulates in long cascaded microring paths. Even when a microring is off resonance, it still contributes some insertion loss, and this loss becomes significant when many rings are placed in series. Therefore, realizing the entire 5-bit input space with a single 16-row cascade would lead to substantial attenuation of the optical signal. By dividing the ROM into four banks and allowing only one bank to be active at a time, the signal interacts with only five cascaded rings during a read operation rather than sixteen. This reduces the overall insertion loss and, at the same time, provides a convenient way to scale the ROM capacity.

Each bank can operate as a digital ROM using a single wavelength at uniform optical power. In this configuration, as shown in Fig.~\ref{fig:scalable_rom}(b), each microring is either electrically connected to a FS block or left unconnected. When a word line selects a specific row, the FS block in that row biases the connected microrings on-resonance (absorbing light, output representing logic `0'), while unconnected microrings remain detuned and do not affect the light (transmitting light, representing logic `1'). The four columns within each bank carry the same wavelength $\lambda_1$, and the optical power at each column is the same. This single-wavelength operation simplifies the optical source requirements and provides deterministic digital readout suitable for lookup table applications requiring binary outputs. We treat the 5-bit input ($-16\ldots+15$) in two parts: the two most significant bits (MSB) select the bank, while the full 5-bit input determines which optical word line (WL) is asserted. The proposed architecture has four banks as shown in Fig.~\ref{fig:scalable_rom}(a), each with four word lines, for a total of 16 word lines, labelled WL$<0>$ through WL$<15>$. The ROM input is first divided at the bank level using the two MSBs, the resulting 2-bit combination selects one of four banks. After that, the required word line is chosen from the four lines available in that bank. Here, each word line stands for two neighbouring decimal inputs rather than a single value. As one example, WL$<0>$ corresponds to -16 and -15, while WL$<1>$ corresponds to -14 and -13. So, taken together, the four WL inside one bank range eight consecutive decimal inputs. Using this mapping, Bank 0, corresponding to MSB = 10, contains WL$<0>$ to WL$<3>$ and represents inputs from -16 to -9. Bank 1, with MSB = 11, contains WL$<4>$ to WL$<7>$ and covers -8 to -1. Bank 2, selected by MSB = 00, uses WL$<8>$ to WL$<11>$ for inputs from 0 to 7. Bank 3, corresponding to MSB = 01, uses WL$<12>$ to WL$<15>$ and covers 8 to 15.

The two MSBs of the input are first used to choose one of the four banks through the selector block in Fig.~2(d). Each bank contains two address rings, A1 and A2, whose PN junctions are preset to store the MSB pattern assigned to that bank. For instance, if a bank corresponds to the address 10, then A1 stores logic 1 and A2 stores logic 0. The wavelengths applied to the selector are also determined by the input MSBs: $\lambda_5$ represents logic 1', while $\lambda_{5B}$ represents logic 0'. Therefore, an input MSB of 10 is encoded by sending $\lambda_5$ and $\lambda_{5B}$ into the selector. In the bank that matches the input address, both rings resonate and absorb the incoming optical signals, so the transmitted power is reduced. This makes the combined power at $P_{EN1}$ and $P_{EN2}$ smaller than the reference power $P_{TH}$. The balanced photodiode therefore outputs a low level, i.e., ground, which enables that bank. For the example considered here, this corresponds to EN$<0>$ = 0 V. At the other banks, the stored address does not match the input address. As a result, one or both rings remain off resonance, and more optical power passes through to the output. Since this power is higher than PTH, the balanced photodiode generates a high output of VDD = 3 V. This gives EN$<1>$ = EN$<2>$ = EN$<3>$ = 3 V, thereby keeping the unmatched banks off. Using this method, every bank sees the same $\lambda_1$ input and word-line signal, but only the matched bank is allowed to respond according to its stored pattern. The selected bank output is then sent to a shared readout layer for digital or analog sensing, as described next.

\subsection{Function Encoding and Reconfigurable Operation}

\begin{figure}[t]
    \centering
    \includegraphics[width=1\linewidth]{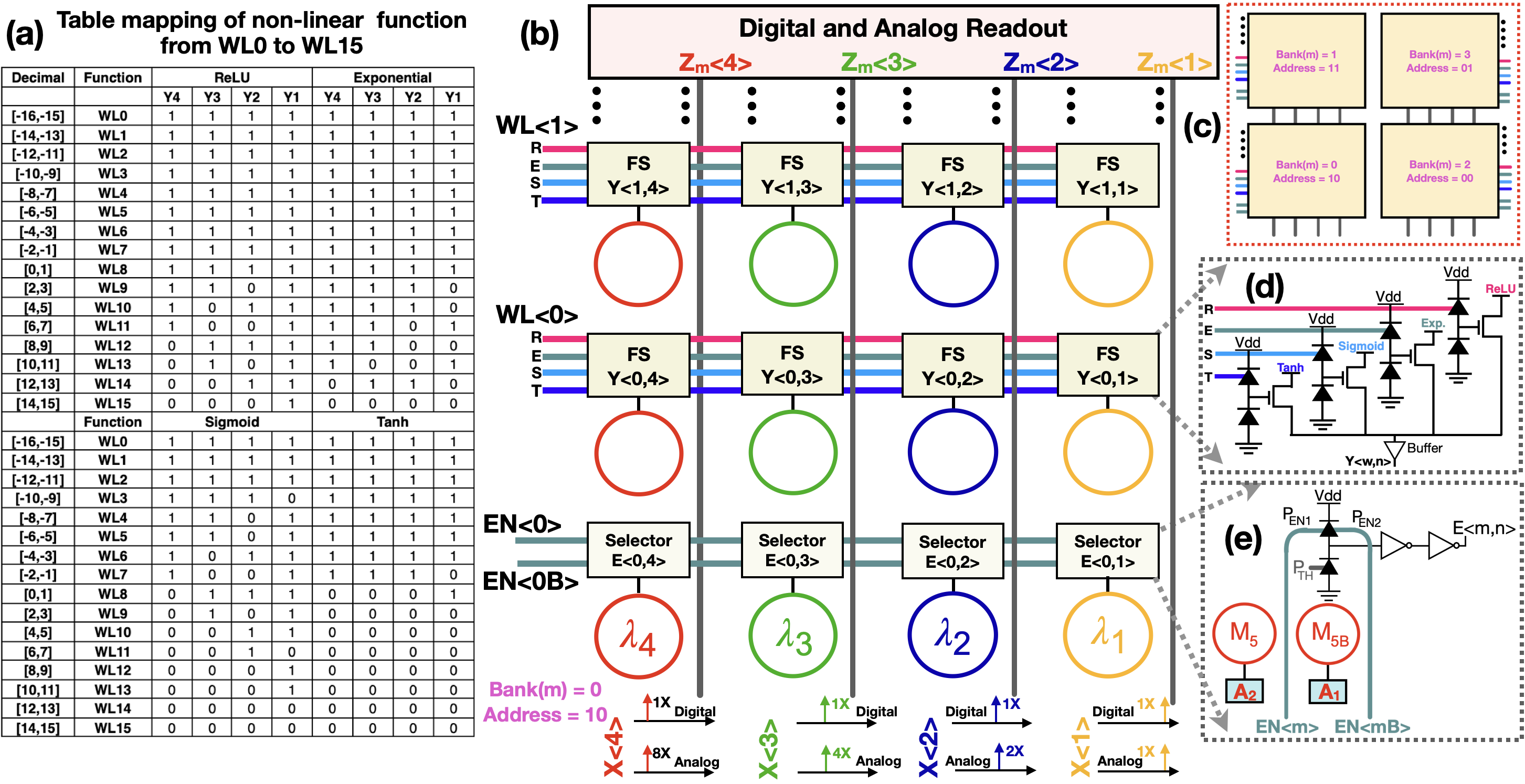}
    \caption{Programmable photonic ROM with REST-encoded function implementation. (a) FS block encoding table showing Vdd (1) and GND (0) connection patterns for ReLU, exponential, sigmoid, and tanh across 16 word lines. (b) Detailed view of Bank 0 showing word line and microring configurations. (c) Four-bank organization with MSB addresses. (d) REST-encoded function-select waveforms. (e) Selector block circuit with address-storing rings.}
    \label{fig:bank_detail}
\end{figure}

Photonic ROM is well-suited for lookup-based operations in applications such as mathematical function evaluation, data encoding, and signal processing. In this work, we further demonstrate that the proposed ROM architecture can implement nonlinear activation functions commonly used in neural network accelerators. In the demonstration, the same microring-based hardware is used to realize four different nonlinear functions, ReLU, exponential, sigmoid, and tanh, by introducing four optical word lines in each row together with a function-selector (FS) block. Reconfigurability is achieved through REST (relu, exponential, sigmoid, tanh)-encoded optical word lines, enabling the ROM to switch between these nonlinear functions without modifying the physical hardware. The nonlinear functions are implemented using the same 16-word-line ROM structure described in the previous section, where each word line is assigned to two adjacent decimal input values ($\Delta x = 2$). Thus, the optical WL serves as the primary row-select signal; for example, when the input is $-16$, WL$<0>$ is asserted, and when the input is $-12$, WL$<2>$ is asserted, as shown in Table~\ref{fig:bank_detail}(a). In this scheme, the continuous function $y = f(x)$ is sampled at 16 evenly spaced points across the 5-bit input range ($x = -16$ to $+15$). Each word line represents two consecutive input values, providing sufficient resolution to approximate the target nonlinear behavior while keeping the hardware compact.
Fig.~\ref{fig:bank_detail}(b) illustrates a detailed view of Bank 0, showing how two word lines (WL$<0>$ and WL$<1>$) are configured with their associated FS blocks and microrings. The complete bank structure, including additional word lines WL$<2>$ and WL$<3>$, follows the same architecture shown in Fig.~\ref{fig:scalable_rom}(b). Fig.~\ref{fig:bank_detail}(c) shows the four-bank organization with their respective MSB addresses. Within the active bank selected by the MSB, a single optical word line is selected based on the incident light input, with the selection controlled by the REST-encoded function shown in Fig.~\ref{fig:bank_detail}(d), thereby targeting a specific row of four microrings.

The integrated diode of each microring is driven by the FS block, which provides an electrical control signal based on the optical word-line input and the selected REST function. As shown in Fig.~\ref{fig:bank_detail}(d), the four REST-encoded optical waveguides are connected to separate balanced photodiodes. When a word line is activated, the optical signal along the selected REST path is converted to an electrical signal that drives the NMOS device's gate. One side of this NMOS is biased with the corresponding voltage listed in Fig.~\ref{fig:bank_detail}(a), while the other side is connected to the microring. As an example, when the ReLU function is chosen, the optical input is applied to the "R" waveguide and detected by its photodiode. If the received optical power is higher than the threshold power $P_{\mathrm{TH}}$, the generated voltage turns the NMOS on, and $Y\langle m,n\rangle$ is driven by the programmed voltage defined in Fig.~\ref{fig:bank_detail}(a). In this operation, an FS output of $Vdd$ tunes the microring into resonance so that the incoming light is blocked, corresponding to logic 0'. On the other hand, an FS output of GND keeps the ring off resonance, allowing the light to propagate and representing logic 1'. The bank selection itself is handled by the selector block in Fig.~\ref{fig:bank_detail}(e), as discussed in the previous section.

\begin{figure}[t]
    \centering
    \includegraphics[width=0.8\linewidth]{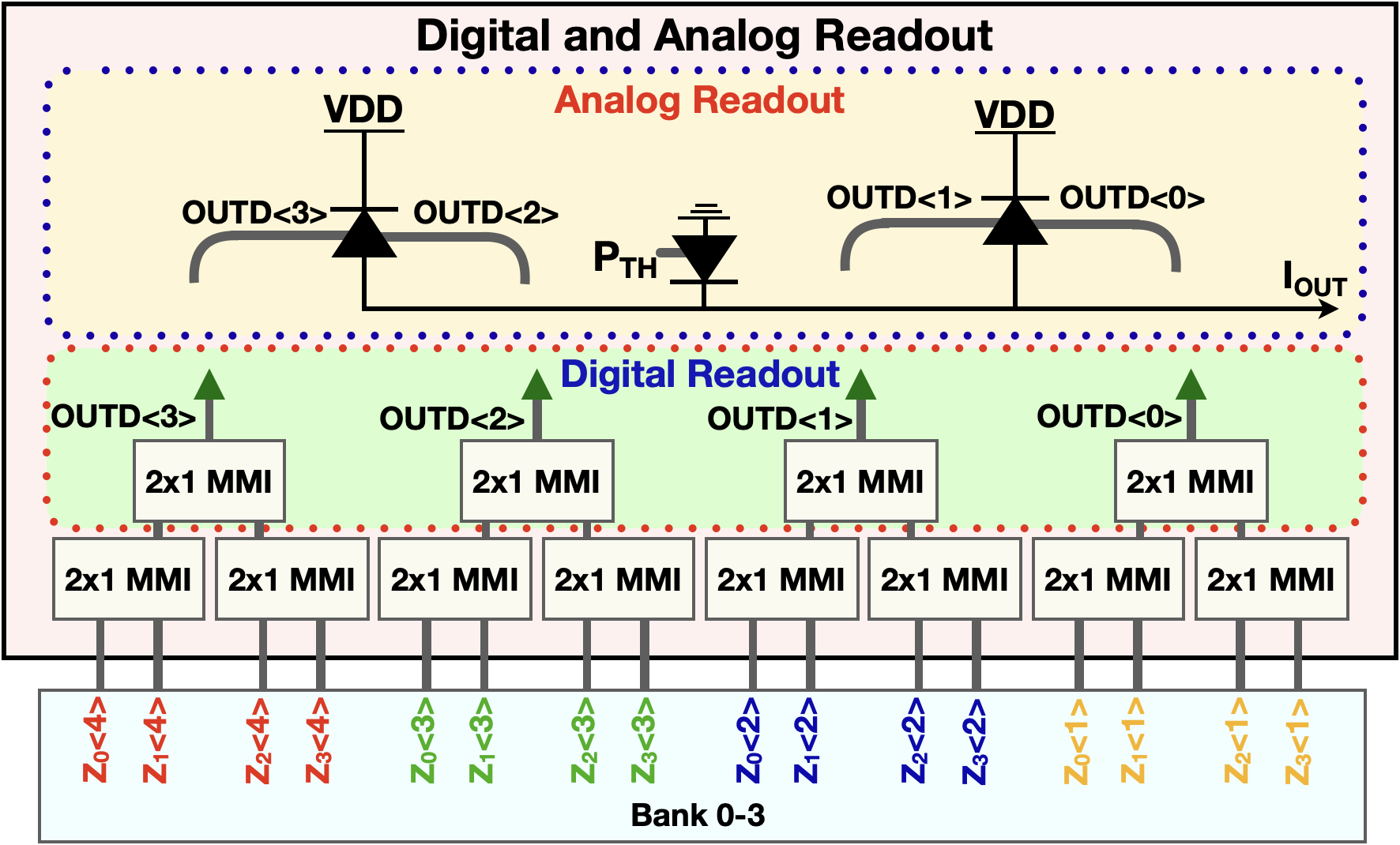}
    \caption{Shared analog and digital readout layer for the proposed banked photonic ROM. Outputs from Banks 0–3 are routed to a common block, where a 2×1 MMI tree provides digital column-wise outputs OUTD$<0>$ – OUTD$<3>$, while an analog branch converts the selected optical power into an electrical current.}
    \label{fig:readout_layer}
\end{figure}

The architecture can operate in two modes, digital and analog, depending on the target application. In the \textbf{digital readout} mode, the ROM uses a single wavelength, $\lambda_1$, with uniform optical input power. The output of each bank is written as $Z_m\langle n \rangle$, where $m$ indicates the bank number and $n$ indicates the column number. Because only one bank is enabled at a time, the outputs can be passed through a cascaded MMI tree, as shown in Fig.~\ref{fig:readout_layer}, to generate the final digital output. For example, if Bank 0 is active and the selected word line corresponds to the stored pattern 1100, then $Z_0\langle 4 \rangle$ and $Z_0\langle 3 \rangle$ have detectable optical power, while $Z_0\langle 2 \rangle$ and $Z_0\langle 1 \rangle$ remain at negligible power. Since the other banks are inactive, their outputs are also close to zero. These signals are then combined through the MMI tree, the final output remains 1100, with OUTD$<3>$ and OUTD$<2>$ representing logic 1' and OUTD$<1>$ and OUTD$<0>$ representing logic 0'. This mode therefore allows the ROM output to be read directly in digital form.

For \textbf{analog readout}, the ROM uses four wavelength channels, $\lambda_1 \ldots \lambda_4$, with binary-weighted optical powers of 1$\times$, 2$\times$, 4$\times$, and 8$\times$, as shown in Fig.~\ref{fig:bank_detail}(b) and established in Fig.~\ref{fig:bitweight_mrr}(c). In this configuration, the bit position directly determines the analog amplitude associated with a given function value. For example, to represent $f(x)=5$, the microrings corresponding to the 1$\times$ and 4$\times$ weights are set to transmit their respective wavelengths, while the 2$\times$ and 8$\times$ channels are suppressed. As a result, the total optical power reaching the output photodetector is proportional to $1+4=5$ units, producing an analog photocurrent ($I_{\mathrm{OUT}}$) that represents the desired function value. This wavelength-multiplexed, power-weighted scheme enables fine-grained analog implementation of nonlinear functions.
Overall, the proposed scalable ROM supports both digital and analog readout modes, making it suitable for applications requiring continuous-valued outputs, such as analog neural-network accelerators, while also enabling digital operation for digital accelerators when needed.

\subsection{Verification of Nonlinear Function}

\begin{figure}[t]
    \centering
    \includegraphics[width=0.9\linewidth]{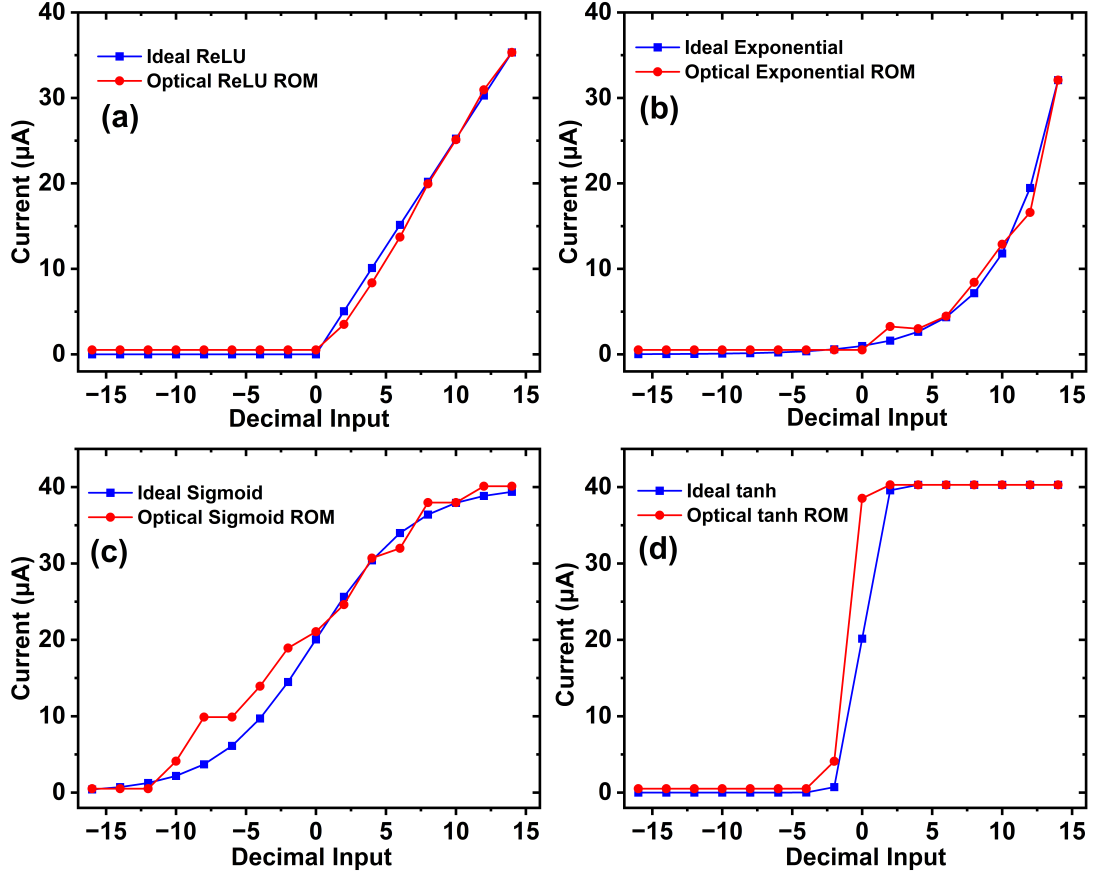}
    \caption{Analog output characteristics of the proposed optical ROM compared with ideal nonlinear activation functions. Blue curves represent the ideal responses, while red curves denote the analog output currents produced by the optical ROM for (a) ReLU, (b) exponential, (c) sigmoid, and (d) tanh.}
    \label{fig:analog_rest}
\end{figure}

\begin{figure}[t]
    \centering
    \includegraphics[width=0.9\linewidth]{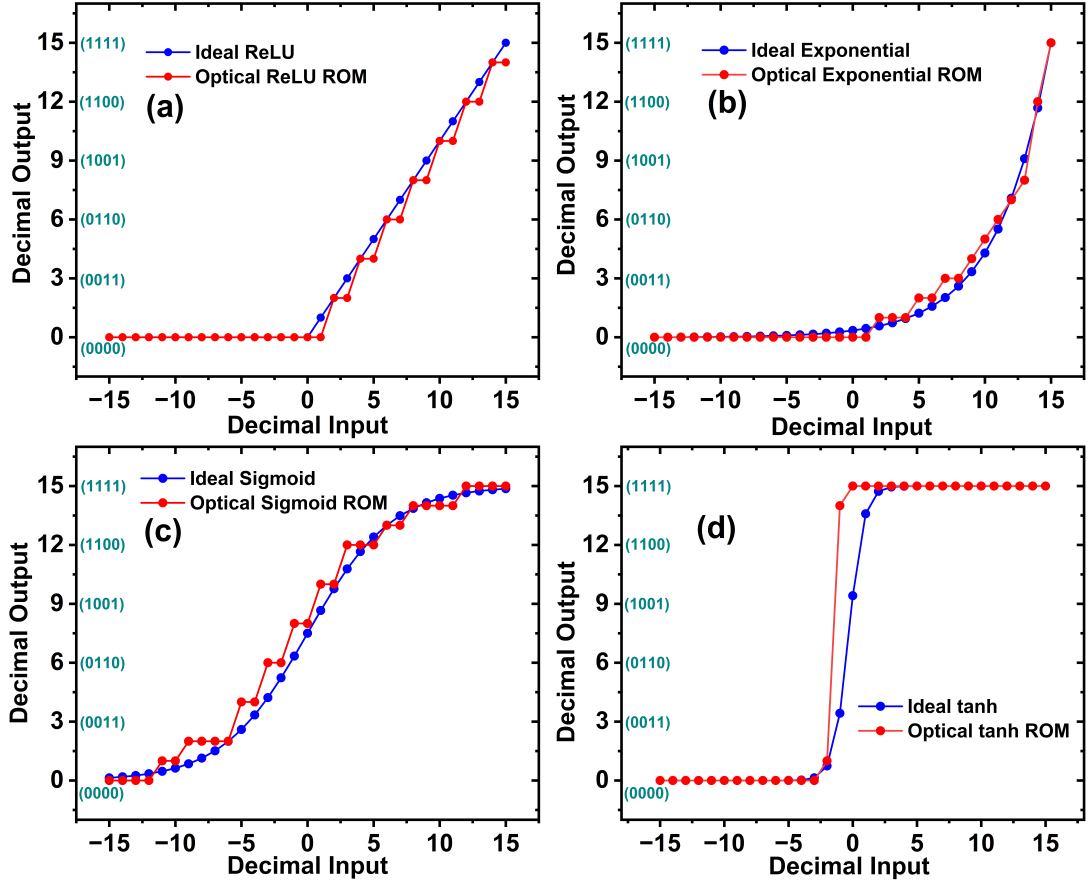}
    \caption{Digitally quantized output characteristics of the proposed optical ROM compared with ideal nonlinear activation functions. Blue curves represent the ideal responses, while red curves denote the decimal output levels generated by the optical ROM for (a) ReLU, (b) exponential, (c) sigmoid, and (d) tanh.}
    \label{fig:digital_rest}
\end{figure}

The proposed design was verified using the GF45SPCLO photonic PDK. Figures~\ref{fig:analog_rest} and~\ref{fig:digital_rest} show the simulated analog and digital responses of the photonic ROM for four activation functions: ReLU, exponential, sigmoid, and tanh. In all cases, the results closely match the ideal function curves, confirming the effectiveness of the REST-encoded word-line approach.

As shown in Fig.~\ref{fig:analog_rest}(a) and Fig.~\ref{fig:digital_rest}(a), the ROM produces a linear rise in output current and digital levels for positive inputs, closely approximating the ideal ReLU curve with near-zero output for negative inputs. The sigmoid function demands finer resolution near the origin, where the gradient is steepest. To capture this behavior, the encoding table in Fig.~\ref{fig:bank_detail}(a) programs the FS blocks so that the central word lines, corresponding to input values near zero, produce finer output transitions, whereas the outer word lines provide the saturated low and high states. As shown in Fig.~\ref{fig:analog_rest}(c) and Fig.~\ref{fig:digital_rest}(c), the resulting sigmoid response agrees well with the theoretical curve, showing a smooth transition from 0 to 40~$\mu$A in the analog case and from 0 to 15 in the digital case across the input range.

The exponential function grows quickly as the input increases, so we use a non-uniform encoding scheme to implement it. More output levels are assigned to values near the origin, where finer representation is needed, while larger input values are mapped more coarsely. The results in Fig.~\ref{fig:analog_rest}(b) and Fig.~\ref{fig:digital_rest}(b) show that the ROM follows the expected exponential behavior. For the tanh function, the current architecture does not support signed output. As a result, the implemented response is not centered at $(0,0)$ as in the ideal tanh function, but instead appears in a shifted non-negative output range with an effective center at $(8,0)$. Near this point, the output changes rapidly, while farther away it gradually saturates to 0 and 15 for digital readout, and to 0 and 40~$\mu$A for analog readout as shown in Fig.~\ref{fig:analog_rest}(d) and Fig.~\ref{fig:digital_rest}(d).

\begin{figure}[t]
    \centering
    \includegraphics[width=0.8\linewidth]{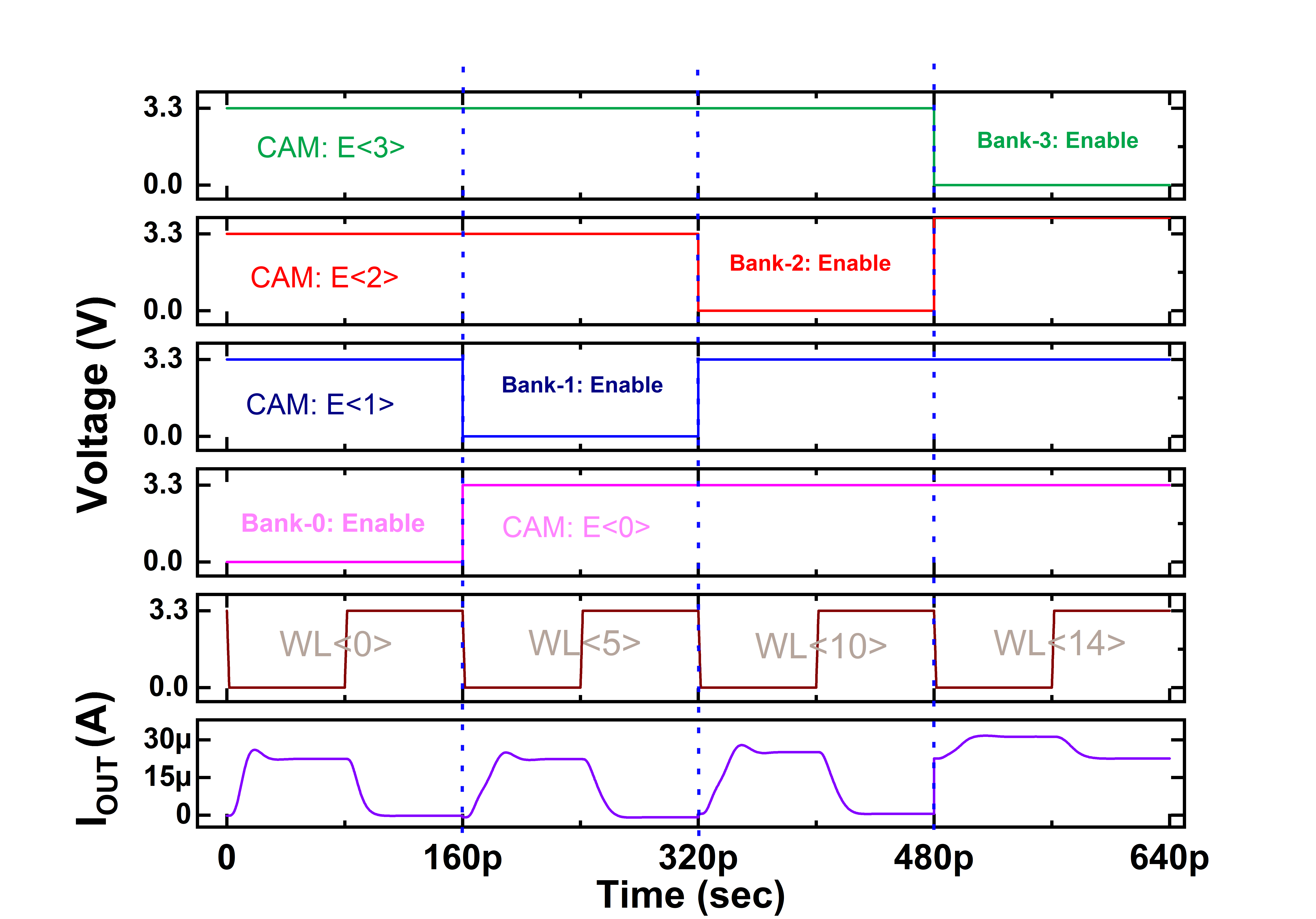}
    \caption{Transient response of the photonic ROM for the ReLU function, showing selector-based bank selection, word-line activation, and output current $I_{\mathrm{out}}$.}
    \label{fig:speed}
\end{figure}

The transient response shown in Fig.~\ref{fig:speed} illustrates the switching dynamics of the proposed photonic ROM for the ReLU function. When the input corresponds to values of $-16$ or $-15$, word line WL$\langle0\rangle$ is activated. In this case, the two MSB are set to 10, enabling Bank~0 via the selector block while all other banks remain disabled at EN = 3~V. The microrings in the disabled banks are forced on-resonance and absorb the incident optical signal, resulting in negligible transmitted optical power, as observed in the time window from 0 to 160~ps. For inputs $-6$ and $-5$, WL$\langle5\rangle$ is activated, enabling Bank~1 while disabling the remaining banks. Subsequently, activation of WL$\langle10\rangle$ enables Bank~2, and activation of WL$\langle14\rangle$ enables Bank~3, with all non-selected banks biased to suppress optical transmission. These bank-selection events are clearly reflected in the transient waveforms. The optical outputs from all banks are routed to the readout layer, where they are converted into a single electrical output current. The output current stabilizes within approximately 80~ps following each word-line transition, indicating a minimum read latency of 80~ps and confirming the ultrafast operation of the proposed photonic ROM architecture.

Power consumption was evaluated using circuit-device simulations that account for both static and dynamic contributions. Under the assumed optical input power and electrical biasing conditions, the total power consumption of the system is estimated to be 2.77~mW, corresponding to an access energy of 259.42~fJ/operation and a device-level energy consumption of 14.74~fJ/device. The architecture achieves reconfigurable nonlinear function evaluation through REST-encoded word line control and hybrid analog/digital readout without requiring external conversion circuitry. The proposed architecture can be extended straightforwardly. New activation functions can be supported by adding their corresponding encoding patterns to the FS-block lookup table, and higher resolution can be achieved by increasing the number of banks or wavelength channels. The operating speed is mainly determined by the accumulated loss through the microring chain. In our case, each ring introduces about 3 dB loss even when it is not resonating, so the optical power drops by nearly half at every stage. After passing through five cascaded rings, the remaining power is only about 1/64 of the original input. Because the final photodetector receives a much weaker optical signal, it takes longer to generate the stable current. Increasing the optical power at the detector would also help improve the operating speed. The current implementation uses 16 optical word lines, each directly activated by the optical input signal. For larger ROM arrays that require more word lines, an optical decoder circuit ~\cite{26sharma2022review,27tian2011demonstration,28daghooghi2018novel} can be employed to decode a smaller number of optical input lines into a larger set of word-line signals, further improving scalability and reducing the number of required optical inputs. This decoder-based approach would enable efficient scaling to higher-bit-depth lookup tables while maintaining the deterministic optical addressing scheme.

\subsection{Performance Benchmarking}

Table~\ref{tab:comparison} compares the OptiLookUp architecture with photonic and electronic ROM implementations. Earlier optical approaches have predominantly relied on semiconductor optical amplifier (SOA)-based architectures to realize lookup or decoder-style functionality. Early demonstrations used reflective SOAs combined with optical decoder structures to implement ROM-like behavior~\cite{19bosu2023design}. Other works reported SOA-based optical ROM using 2-to-4 decoder configurations composed of optical power splitters, combiners, and encoders, enabling all-optical address decoding and data retrieval~\cite{17mukherjee2011all}. Simplified one-level all-optical ROM architectures using SOAs have also been proposed to reduce structural complexity while maintaining high-speed operation~\cite{18jung2009design}. In addition, microring-resonator-based optical ROMs implemented on silicon nitride (Si$_3$N$_4$) platforms have demonstrated moderate-scale memory arrays, such as an $8\times6$ ROM, highlighting the feasibility of resonance-based optical storage~\cite{20saharia2022proposed}. While these prior approaches establish the viability of optical ROM concepts, they typically incur higher energy consumption, increased footprint, or limited scalability, motivating the development of more compact, energy-efficient, and reconfigurable photonic ROM architectures. In most prior works, photonic ROM functionality was introduced to demonstrate fixed lookup or decoder operations rather than explicit nonlinear function evaluation. Nevertheless, the inherent lookup-based nature of ROM architectures is well-suited to implementing nonlinear functions that require predefined input-output mappings. This functional alignment motivates the use of photonic ROMs as efficient hardware for the realisation of nonlinear functions.

\begin{table*}[t]
\centering
\caption{Comparison of photonic and electronic lookup-based implementations relevant to nonlinear function realization.}
\label{tab:comparison}
\begin{tabular}{l p{4cm} p{2cm} p{4cm}}
\hline
\textbf{Device / Architecture} & \textbf{Energy/Operation} & \textbf{Latency} & \textbf{Demonstrated Functionality} \\ 
\hline
Reflective SOA ROM~\cite{19bosu2023design} 
& 40–400 fJ & -- 
& \textit{Fixed optical lookup, decoder-based ROM} \\ 

SOA-based decoder ROM (2--4)~\cite{17mukherjee2011all} 
& 1.4–2.3 pJ & 40~ps 
& \textit{All-optical address decoding and ROM readout} \\ 

Simplified SOA-based ROM~\cite{18jung2009design} 
& 0.8–2 pJ & 200~ps 
& \textit{One-level all-optical ROM, lookup functionality} \\ 

Si$_3$N$_4$ microring ROM ($8\times6$)~\cite{20saharia2022proposed} 
& -- & --
& \textit{Resonance-based optical ROM, wavelength-addressed lookup} \\ 

MRAM ROM~\cite{25fong2015embedding} 
& 128 fJ & 1–1.2~ns 
& \textit{sin(x), log(x)} \\ 

\textbf{OptiLookUp (This Work)} 
& \textbf{259.42 fJ} 
& \textbf{80 ps} 
& \textit{ReLU, Exponential, Sigmoid, Tanh} \\ 
\hline
\end{tabular}
\end{table*}

OptiLookUp introduces a reconfigurable photonic ROM architecture in which nonlinear functions are stored as lookup entries in banked microring arrays. The proposed design offers several practical advantages. First, the banked ROM structure improves scalability while reducing the insertion loss that would otherwise accumulate in long cascaded microring paths. Second, it is naturally compatible with accelerator architectures, where activation functions are commonly implemented through lookup tables. Third, the architecture supports high-speed operation with strong energy efficiency. Finally, the architecture is intrinsically reconfigurable. Beyond the current REST functions, additional nonlinear functions can be implemented by adding their corresponding optical waveguides and assigning the required FS-block encoding patterns. 

Specifically, OptiLookUp achieves an access energy of 259.42~fJ/operation and a device-level energy consumption of 14.74~fJ/device, with a latency of approximately 80~ps (corresponding to 12.5~GHz operation), thereby positioning it among the fastest optical ROM implementations reported to date while maintaining competitive energy efficiency. OptiLookUp thus effectively bridges the gap between electronic and photonic computing: it combines the precision and determinism of digital LUTs with the ultra-fast parallelism of photonics, enabling nonlinear activation functions to be executed at speeds proportional to modern optical interconnects while retaining the flexibility to adapt to different application requirements.

\section*{Conclusion}

We present a reconfigurable photonic read-only memory architecture based on banked microring arrays with optical selector-based addressing. The banked organization limits optical path length to fewer microrings per access, substantially reducing insertion loss while enabling scalable expansion. Simulations using the GlobalFoundries 45SPCLO photonic PDK demonstrate operation at 12.5~GHz with 259.42~fJ/operation energy and 80~ps latency, achieving competitive performance with prior optical ROM implementations. To validate the architecture's flexibility, we implement four nonlinear activation functions (ReLU, sigmoid, tanh, exponential) as representative lookup-based mappings. The simulated outputs closely match the theoretical curves, demonstrating reconfigurable function storage through programmable FS block encoding patterns. This reconfigurability, combined with low latency and hybrid analog/digital readout capability, positions the photonic ROM as a versatile building block for applications ranging from photonic neural networks to optical signal processing and programmable photonic systems.

\begin{acknowledgement}

This work was supported by the Defense Advanced Research Projects Agency (DARPA) under Grant No. N660012424003, and in part by the DoD ASPEN program and the NSF under Award No. CCF-2319617.

\end{acknowledgement}


\bibliography{achemso-demo}

\end{document}